\documentclass[fdp,a4paper,fleqn%
]{w-art}
\usepackage{times,cite,w-thm}
\usepackage{url}
\usepackage{amssymb}
\theoremstyle{plain}
\usepackage[]{graphicx}

\newcommand{\ud}{\mathrm{d}\,}

\begin{document}
\DOIsuffix{theDOIsuffix}
\Volume{55}
\Month{01}
\Year{2007}
\pagespan{1}{}
\Receiveddate{}
\Reviseddate{}
\Accepteddate{}
\Dateposted{}
\title[A Mellin Space Program for $W^{\pm}$ and $Z^0$ Production at NNLO]{A Mellin Space Program for $W^{\pm}$ and $Z^0$ Production at NNLO}

\author[Petra Kova\v{c}\'{i}kov\'{a}]{Petra Kova\v{c}\'{i}kov\'{a} %
  \footnote{Corresponding author\quad E-mail:~\textsf{petra.kovacikova@desy.de},
  Phone: +00\,493\,376\,277\,193 }}
\address{Deutsches Elektronen-Synchrotron, Platanenallee 6, D-15738 Zeuthen, Germany}
\begin{abstract}
	We present a program for the evaluation of full unpolarized cross sections for the $W^{\pm}$ and
	$Z^0$ production in the narrow width approximation at NNLO in perturbative QCD using Mellin
	space techniques.
\end{abstract}
\maketitle                   

\section{Introduction}

The Drell-Yan process, originally described in the context of the parton model~\cite{Drell:1970wh},
concerns the production of a lepton pair of large invariant mass in hadron-hadron collisions.
With the increase of the centre of mass energy at particle accelerators, the Drell-Yan process led to
the discovery of $W^{\pm}$ and $Z^{0}$ bosons at UA1 and UA2 experiments~\cite{Albajar:1987yz, Alitti:1991dm}. Since then 
the properties of  massive vector bosons have been studied in great detail.
At present the production of $W^{\pm}$ and $Z^{0}$  provides an important benchmark for the LHC and a test of the Standard Model (SM) in  a new range of
centre of mass energies~\cite{Alekhin:2010dd}.

As guaranteed by the factorisation theorem~\cite{Collins:1985ue}, one can separate the physics
of soft energy scales from the  physics at hard energy scales where perturbation theory applies. The
higher order QCD corrections to the Drell-Yan process have been calculated up
to next-to-next-to-leading order (NNLO), see~\cite{Altarelli:1979ub, Hamberg:1990np, Hamberg2002403} and references therein.
The full cross section is obtained as a convolution with the parton distribution functions (PDFs) that encode the non-perturbative information. 

In this paper, we present a program for evaluation of the full inclusive cross section for $W^{\pm}$ and
$Z^{0}$ production in a fast and accurate way using a Mellin space approach. After a brief description of the basic
ingredients of the calculation we give formulae for the Mellin transforms. 
We then present a comparison with the code ZWPROD~\cite{Hamberg:1990np,Hamberg2002403} and discuss possible applications and
extensions within this framework.

\section{Formalism}

We consider the inclusive production of a single vector boson $V=W^+, W^-$ or $Z^0$ in hadron-hadron collision with a centre
of mass energy $s$ which subsequently decays into a lepton pair of an invariant mass $Q^2$.
The decay of the vector boson is treated within the narrow width approximation which  
replaces the propagator by a delta function such that $Q^2 = M^2_V$. We consider massless quarks. The
cross section for this process can be expressed as 
\begin{equation}
	\sigma^{h_1 h_2\to V \to l_1l_2} (s)
	= x \sigma^{V \to l_1l_2} W^V(x,Q^2), \qquad x = Q^2/s, \label{sigma}
\end{equation}
where $\sigma^{V \to l_1l_2 }$ represents the kinematically independent part of the Born level subprocess 
$q\bar{q} \to V \to l_1l_2 $ (the point-like cross section)
 multiplied by the appropriate branching
ratio. The exact form of the point like cross section can be found in Ref.~\cite{Hamberg:1990np},
formulae (A.10) and (A.11). \footnote{There is an extra factor of 2 in the
denominator of the formula (A.11), corrected by~\cite{Anastasiou:2003ds}} 

The structure  function $W^V(x,Q^2)$ is written as a convolution of two
parton distribution functions $f_{a}(\mu_f^2)$ and $f_{b}(\mu_f^2)$ and a hard scattering cross
section represented by coefficient functions $\Delta_{ab}$,
\begin{eqnarray} 
	W^V(x,Q^2) &=& \sum_{a,b = q,\bar{q},g} C^V_{a,b} \Big[f_{a}(\mu_f^2)\otimes
	f_{b}(\mu_f^2) \otimes \Delta_{ab}(Q^2,\mu_f^2,\mu_r^2)\Big](x).  \label{sf} 
\end{eqnarray}
The perturbative coefficients are known up to NNLO~\cite{Hamberg:1990np,Hamberg2002403}, 
\begin{equation}
	\Delta^{\textrm{N}k\textrm{LO}}_{ab}(x,Q^2,\mu_f^2,\mu_r^2) = \sum_{n=0}^k
	\frac{\alpha^{k}_s(\mu_r^2)}{4\pi}\Delta^{(k)}_{ab}(x,Q^2,\mu_f^2,\mu_r^2). \label{delta}
\end{equation}
The factor $C^V_{a,b}$ in Eq.(\ref{sf}) contains information about couplings of vector bosons to partons $a$ and $b$. 
For the detailed form of the Eq.(\ref{sf}) we refer the reader to the paper
of Hamberg, Matsuura and van Neerven~\cite{Hamberg:1990np} whose notation we follow closely
\footnote{Several typos appearing in Ref.~\cite{Hamberg:1990np} have been pointed out in~\cite{Anastasiou:2003ds}}. 
The convolution sign represents an integral 
\begin{eqnarray}
	&{}&(f_1\otimes f_2 \otimes \dots
	\otimes f_k)(x)\nonumber\\ &{}& = \int_0^1\ud x_1\, \int_0^1 \ud x_2\,\dots \int_0^1 \ud
	x_k\, \delta(x - x_1 x_2\dots x_k) f_1(x_1)f_2(x_2)\dots f(x_k). \label{convolution}
\end{eqnarray}
In principle one can perform the integrals in Eq.(\ref{sf}) directly however, the problem is much
better addressed after transforming  to Mellin space, 
\begin{eqnarray}
        f(N) &=& \int_0^1\ud x\, x^{N-1}f(x),
	\label{MellinTransform} 
\end{eqnarray}
This transformation turns the integrals  in Eq.(\ref{sf}) into ordinary products such that the structure function reads
\begin{eqnarray}
	W^V(N,Q^2)
	&=& \sum_{a,b = q,\bar{q},g} C^V_{a,b} f_{a}(N,Q^2) f_{b}(N,Q^2)  \Delta_{ab}(N,Q^2)\qquad
	\mu_f=\mu_r=Q^2,
        \label{nsf}
\end{eqnarray}
and therefore it is possible to evaluate it in  a fast and efficient way. The formula for the inverse
Mellin transform defines how to recover the original momentum space result, 
\begin{equation}
         W^V(x,Q^2) =  \frac{1}{2\pi i}\int_{c-i\infty}^{c+i\infty}\ud N\,x^{-N} W^V(N,Q^2),
	\label{InverseMellinTransform} 
\end{equation}
where $c$ represents a point on the real axis such that all poles $N_i$ in the function $W(N,Q^2)$ lie to the left from $c$.
Further on, we will refer to functions in Mellin space as $N$ space functions and functions from momentum space as $x$ space functions.

\section{Implementation}

The main ingredients of the calculation are the coefficient functions up to NNLO and the parton
distribution functions in Mellin space in terms of a complex variable $N$. The condition
$N\in\mathbb{C}$  is required for the numerical evaluation of the inversion formula
(\ref{InverseMellinTransform}).
For this we adopted the technique implemented in QCD-PEGASUS~\cite{Vogt:2004ns}. 
The complex integral (\ref{InverseMellinTransform}) is rewritten in terms of an integral over a real variable $z$ 
\begin{equation}
W(x, Q^2) = \frac{1}{\pi}\int_0^\infty \ud z\,
\textrm{Im}[e^{i\varphi}x^{-c-z e^{i\varphi}}W(N,Q^2)] \qquad N = c + z\exp^{i\phi} \in \mathbb{C}
  \label{InverseMellinTransformNum}
\end{equation}
and evaluated using Gaussian quadratures. The parameter  $\phi > \pi/2$  represents the angle with
respect to the positive real axis. Since the rightmost pole of the structure function is $N_{\textrm{max}}= 1$, we chose $c=1.5$. 
These values as well as the maximum value of the integration variable $z$ are flexible and can be modified by the user in the main
program if desired.
For a more detailed description of the shape of the integration contour we refer to the QCD-PEGASUS manual~\cite{Vogt:2004ns}.

The coefficient functions in  $N$ space were published previously in Ref.~\cite{Blumlein:2005im}, including corrections to the previous literature.
The corresponding {\tt FORTRAN} code is {\tt DY.f} used together with {\tt ANCONT}\cite{Blumlein:2000hw}.
We performed the Mellin transforms starting from the $x$ space expressions~\cite{Ravindran:unpublished} using the \texttt{harmpol} package~\cite{Remiddi:1999ew}. 
The results can be expressed mostly in terms of complex-valued simple harmonic sums~\cite{Vermaseren:1998uu,
Blumlein:1998if} and several more complicated ones which we approximated by using the minimax
method\footnote{We used the MINIMAX routine implemented in Maple}
worked out in detail in~\cite{Blumlein:2000hw} previously.\footnote{Exact expressions were given in \cite{Blumlein:2009ta}.}
The absolute accuracy of our approximation is better than $10^{-9}$ over the whole kinematic range.

At the moment there are two options for the input parton distribution functions in $N$ space. A toy input corresponds to the one used for the 2001/2 benchmark tables~\cite{Giele:2002hx} and is used for comparisons with ZWPROD~\cite{Hamberg:1990np,Hamberg2002403}   assuming no evolution of PDFs. The general form reads
\begin{equation}
	xf_{i,\textrm{toy}}(x,\mu_0^2)  = n x^a(1-x)^b, \qquad i = q,\bar{q},g, \qquad n,a,b \in
	\mathbb{R},	
	\label{toypdfx}
\end{equation}	
which is in Mellin space represented by an Euler beta function
\begin{equation}
	f_{i,\textrm{toy}}(N,\mu_0^2)  = n\beta(a+N,b+1).
	\label{toypdfn}
\end{equation}
The second option for the PDF input is using the FORTRAN code QCD-PEGASUS~\cite{Vogt:2004ns} which
can be linked to our program. 

\begin{vchfigure}[h]
\includegraphics[angle=-90,width=0.7 \textwidth]{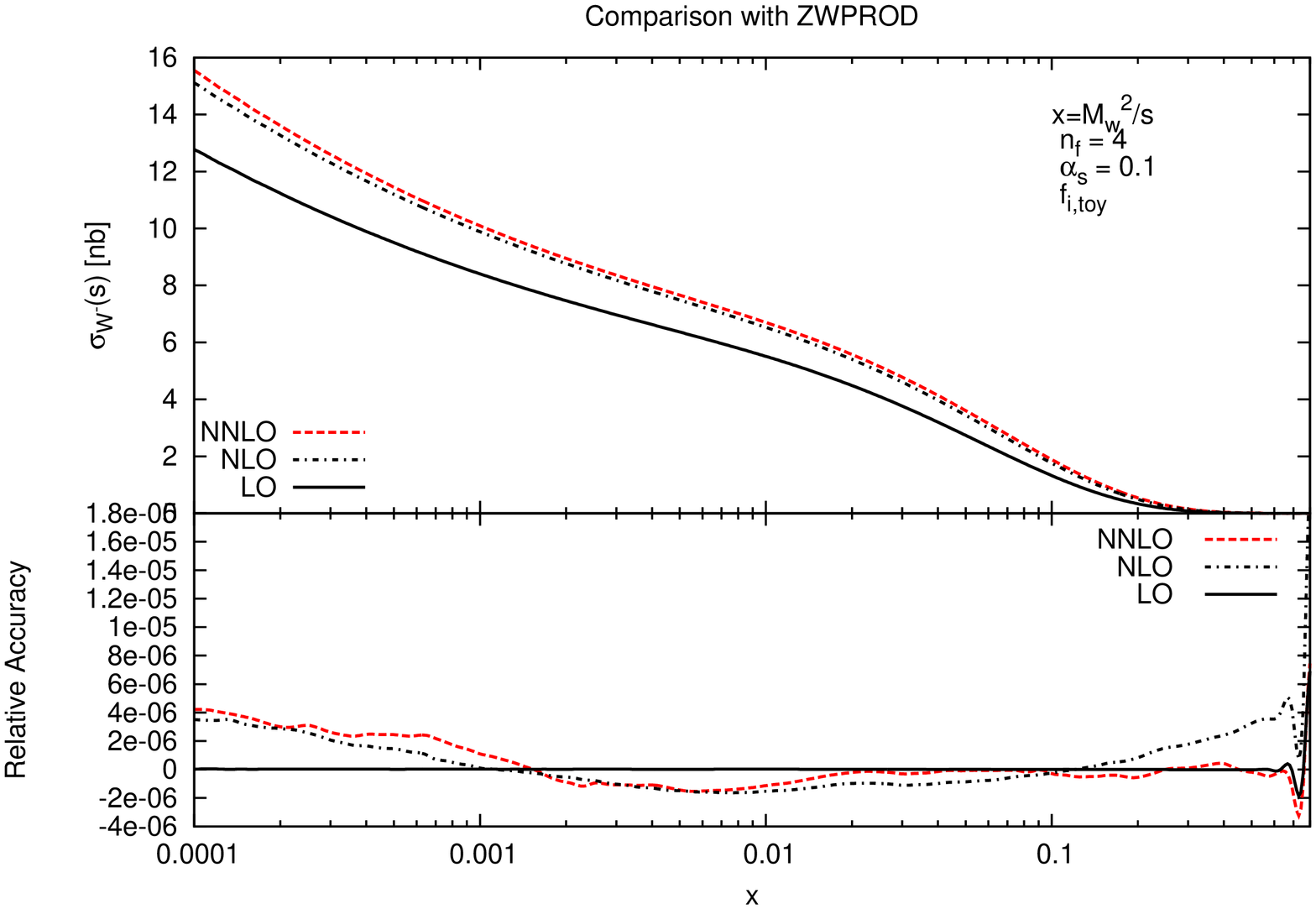}
\vchcaption{Cross section for $W^-$ production up to NNLO in the narrow width approximation using the toy parton distribution functions and a fixed value of the strong coupling constant. Upper part: The full cross section. Lower part: Relative accuracy with respect to the ZWPROD.}
\label{figure1}
\end{vchfigure}

\section{Results And Outlook}

There are several programs on the market using the standard momentum space evaluation~\cite{Catani:2007vq,Catani:2009sm,Gavin:2010az} which can provide a cross-check 
for our $N$ space calculation. 
We performed comparisons of the full cross sections with a program ZWPROD written by the authors of the original calculation of the NNLO  Drell-Yan coefficient functions~\cite{Hamberg:1990np,Hamberg2002403}. The Fig.~(\ref{figure1}) shows a
comparison for the $W^-$ cross section using toy input for
PDFs corresponding to the Eq.(\ref{toypdfn}) with no evolution and a fixed value of the coupling constant $\alpha_s = 0.1$. The relative accuracy is
better than $6\times10^{-6}$ in the relevant kinematical range  $x \in (10^{-4},0.8)$. 
As an intermediate check, we compared the Mellin inversion of $N$ space coefficient functions against the $x$ space expressions using a program of Gehrmann and Remiddi~\cite{Gehrmann:2001pz} for the numerical evaluation of harmonic polylogarithms.
The framework presented here is suitable for a further implementation of those cross sections where $N$
space coefficient functions are also available, like Higgs production and deep inelastic
scattering (DIS)~\cite{Blumlein:2005im, vanNeerven:2000uj,Vogt:2006bt,Moch:2004xu,Vermaseren:2005qc}.
The setup is well suited for merging the program with  threshold resummation calculations which
are typically performed in Mellin space (see e.g.~\cite{Moch:2005ba}).
For the extraction of PDFs from $W^{\pm}$ and $Z$ production it would be desirable to have an access to the rapidity distributions
in which case one will need to apply double Mellin transforms of two variables $N_1$ and $N_2$ however, this is a subject to further study.
On the side of PDFs we aim for a direct interface to the LHAPDF grids~\cite{Whalley:2005nh} which will allow  the user to freely choose any particular PDF set provided within this framework.  
Recent results~\cite{bhkm} on $N$ space input parametrizations also allow for more flexible input PDF parametrisations in QCD-PEGASUS.
Further improvements with respect to the speed of the code are foreseen and together with an upgrade on the input PDFs this code can become a tool for PDF fits, where fast and accurate evaluations of cross sections are needed.
The current version of the c++ code can be downloaded  from  \url{http://www-zeuthen.desy.de/~kpetra/sbp}.

\begin{acknowledgement}
	I would like to thank my advisor Sven-Olaf Moch for continuous support and advice
	throughout the course of this work and for careful reading of the manuscript. I am grateful to Prof. Johannes Bl\"umlein for useful discussions and comments. 
	This work was supported by the Marie-Curie Research Training Networks MRTN-CT-2006-035505 HEPTOOLS.
\end{acknowledgement}

 \bibliographystyle{fdp}
	
\providecommand{\WileyBibTextsc}{}
\let\textsc\WileyBibTextsc
\providecommand{\othercit}{}
\providecommand{\jr}[1]{#1}
\providecommand{\etal}{~et~al.}

 \bibliography{bibliography}
\end{document}